\documentclass[prl,twocolumn,superscriptaddress,preprintnumbers,amsmath,amssymb]{revtex4}
\usepackage{graphicx}
\usepackage{amsmath}
\usepackage{latexsym}
\usepackage{dcolumn}
\usepackage{bm}
\usepackage{float}
\usepackage{graphicx,here}
\usepackage{color}

\emergencystretch=\maxdimen
\begin{document}

\title{\bf Experimental observation of Dirac cones in artificial graphene lattices}

\affiliation{Institute of Physics, Chinese Academy of Sciences, Beijing 100190, China}
\affiliation{Hiroshima Synchrotron Radiation Center, Hiroshima University, 2-313 Kagamiyama, Higashi-Hiroshima 739-0046, Japan}
\affiliation{School of Physical Sciences, University of Chinese Academy of Sciences, Beijing 100049, China}
\affiliation{Songshan Lake Materials Laboratory, Dongguan, Guangdong 523808, China}

\author{Shaosheng Yue}
\affiliation{Institute of Physics, Chinese Academy of Sciences, Beijing 100190, China}
\affiliation{School of Physical Sciences, University of Chinese Academy of Sciences, Beijing 100049, China}
\author{Hui Zhou}
\affiliation{Institute of Physics, Chinese Academy of Sciences, Beijing 100190, China}
\affiliation{School of Physical Sciences, University of Chinese Academy of Sciences, Beijing 100049, China}
\author{Daiyu Geng}
\affiliation{Institute of Physics, Chinese Academy of Sciences, Beijing 100190, China}
\affiliation{School of Physical Sciences, University of Chinese Academy of Sciences, Beijing 100049, China}
\author{Zhenyu Sun}
\affiliation{Institute of Physics, Chinese Academy of Sciences, Beijing 100190, China}
\affiliation{School of Physical Sciences, University of Chinese Academy of Sciences, Beijing 100049, China}
\author{Masashi Arita}
\affiliation{Hiroshima Synchrotron Radiation Center, Hiroshima University, 2-313 Kagamiyama, Higashi-Hiroshima 739-0046, Japan}
\author{Kenya Shimada}
\affiliation{Hiroshima Synchrotron Radiation Center, Hiroshima University, 2-313 Kagamiyama, Higashi-Hiroshima 739-0046, Japan}
\author{Peng Cheng}
\affiliation{Institute of Physics, Chinese Academy of Sciences, Beijing 100190, China}
\affiliation{School of Physical Sciences, University of Chinese Academy of Sciences, Beijing 100049, China}
\author{Lan Chen}
\affiliation{Institute of Physics, Chinese Academy of Sciences, Beijing 100190, China}
\affiliation{School of Physical Sciences, University of Chinese Academy of Sciences, Beijing 100049, China}
\affiliation{Songshan Lake Materials Laboratory, Dongguan, Guangdong 523808, China}
\author{Sheng Meng}\thanks{smeng@iphy.ac.cn}
\affiliation{Institute of Physics, Chinese Academy of Sciences, Beijing 100190, China}
\affiliation{School of Physical Sciences, University of Chinese Academy of Sciences, Beijing 100049, China}
\author{Kehui Wu}\thanks{khwu@iphy.ac.cn}
\affiliation{Institute of Physics, Chinese Academy of Sciences, Beijing 100190, China}
\affiliation{School of Physical Sciences, University of Chinese Academy of Sciences, Beijing 100049, China}
\affiliation{Songshan Lake Materials Laboratory, Dongguan, Guangdong 523808, China}
\author{Baojie Feng}\thanks{bjfeng@iphy.ac.cn}
\affiliation{Institute of Physics, Chinese Academy of Sciences, Beijing 100190, China}
\affiliation{School of Physical Sciences, University of Chinese Academy of Sciences, Beijing 100049, China}

\date{\today}

\clearpage

\begin{abstract}
Artificial lattices provide a tunable platform to realize exotic quantum devices. A well-known example is artificial graphene (AG), in which electrons are confined in honeycomb lattices and behave as massless Dirac fermions. Recently, AG systems have been constructed by manipulating molecules using scanning tunnelling microscope tips, but the nanoscale size typical for these constructed systems are impossible for practical device applications and insufficient for direct investigation of the electronic structures using angle-resolved photoemission spectroscopy (ARPES). Here, we demonstrate the synthesis of macroscopic AG by self-assembly of C$_{60}$ molecules on metal surfaces. Our theoretical calculations and ARPES measurements directly confirm the existence of Dirac cones at the $K$ ($K^\prime$) points of the Brillouin zone (BZ), in analogy to natural graphene. These results will stimulate ongoing efforts to explore the exotic properties in artificial lattices and provide an important step forward in the realization of novel molecular quantum devices.
\end{abstract}

\maketitle

Graphene is a single layer of carbon atoms with a honeycomb lattice, and it has been intensively studied in the past decade \cite{GeimAG2007,NetoAH2009}. In the proximity of the Fermi level, the electrons in graphene behave as massless Dirac fermions. This is the origin of graphene's various exotic properties, such as half-integer quantum Hall effects \cite{NovoselovKS2005,ZhangY2005} and the Klein paradox \cite{KatsnelsonMI2006,YoungAF2009,GutierrezC2016}. An alternative way to realize the exotic properties of graphene is by confining the electrons of a two-dimensional electron gas (2DEG) to an equivalent honeycomb lattice, called artificial graphene (AG) \cite{GomesKK2012,WangS2014,NantohM2017}. Because of the confinement, there exists a Dirac cone at each $K$ point of the BZ. As a result, electrons of the otherwise 2DEG behave as massless Dirac fermions, which is analogous to the case of natural graphene. In addition to possessing the novel properties of graphene, artificial lattices possess various tunable parameters, thereby providing an ideal platform for the simulation of quantum behaviours in two-dimensional (2D) Dirac materials \cite{SinghaA2011,TarruellL2012,PoliniM2013,KempkesSN2019,KempkesSN2019'}.

Some molecules, such as carbon monoxide and coronene, can serve as potential barriers through which the electrons of the 2DEG are forbidden to travel. Therefore, a hexagonally patterned molecular lattice can confine the 2DEG electrons into an equivalent honeycomb lattice and lead to the realization of AG \cite{GomesKK2012,WangS2014}, as illustrated in Figs. 1(a) and 1(b). In 2012, Gomes {\it et al.} constructed the first AG by atomic manipulation of carbon monoxide molecules on Cu(111) and found experimental evidence of massless Dirac fermions by scanning tunnelling spectroscopy \cite{GomesKK2012}. In addition to the scheme presented by molecular AG, AG has also been realized in nanopatterned GaAs quantum wells \cite{GibertiniM2009,WangS2016,WangS2018}. Unlike unpatterned GaAs quantum wells, resonant inelastic light-scattering spectra of the AG system revealed low-lying transitions that might arise from the Dirac bands \cite{WangS2018}. The high energy and momentum resolution of ARPES makes it a powerful technique to directly study the electronic structures of materials. However, direct experimental observation of the Dirac cones in AG using ARPES has not yet been reported.

\begin{figure*}[htb]
\centering
\includegraphics[width=15 cm]{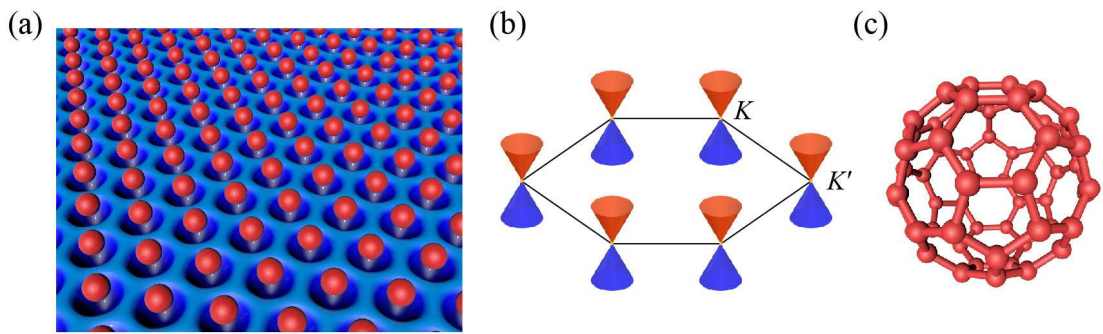}
\caption{{\bf Schematic of artificial graphene (AG).} (a) Molecular AG system, including molecules that act as potential barriers for electrons (red balls) and the electron density profile (blue surfaces). Because of the presence of hexagonally-patterned molecules, electrons of the otherwise two-dimensional electron gas are confined in a honeycomb lattice. (b) Band structure of AG. A Dirac cone exists at the $K$ ($K^\prime$) point of the Brillouin zone (BZ), as an analogy to graphene. (c) Structure model of a C$_{60}$ molecule. The nearly isotropic shape of C$_{60}$ makes it an ideal choice for the construction of AG.}
\end{figure*}

Here, we constructed macroscopic AG systems by growing monolayer C$_{60}$ molecules on the (111)-terminated surfaces of noble metals, which enabled ARPES studies of the electronic band structure. Our low-energy electron diffraction (LEED) and ARPES measurements show that these AG systems are homogeneous. Therefore, their size is only limited by the scale of the substrates. The Dirac cones at the $K$ ($K^\prime$) points were directly observed by our ARPES measurements. In addition, we performed model calculations on these AG systems, whose results fully supported our experimental observations.

The lattice constant of nanopatterned structures is typically tens of hundreds of nanometres, which is several orders of magnitude larger than that of conventional single crystals \cite{GibertiniM2009,WangS2016,WangS2018}. Such a large lattice constant results in a very small BZ that is beyond the resolution of conventional ARPES facilities. The molecular AG system is ideal for the investigation of electronic structures because of its moderate lattice constants. However, the size of conventional AG systems constructed by atomic/molecular manipulation is limited to the nanoscale, which is typically insufficient for practical device applications and for electronic band structure investigation. A promising route to realize macroscopic AG is by preparing a molecular self-assembled monolayer on a metal substrate; where such supramolecular architectures can be well ordered across the entire substrate surface. A particular interesting molecule is C$_{60}$, which has a nearly-isotropic spherical shape, as shown in Fig. 1(c). Previous works have shown that C$_{60}$ molecules can form hexagonal structures on various noble metal surfaces, including Cu(111), Au(111), and Ag(111). In these systems, the electrons of the 2DEG of the metal surfaces are confined in an equivalent honeycomb lattice and are thus expected to behave as massless Dirac fermions.

First, we studied C$_{60}$ monolayers on Cu(111). The C$_{60}$ molecules form a hexagonal structure with a 4$\times$4 superstructure with respect to the 1$\times$1 lattice of Cu(111) \cite{TamaiA2008,PaiWW2010}. The low-energy electron diffraction (LEED) patterns of the C$_{60}$ monolayer on Cu(111) are shown in Fig. S1 \cite{SM}. The BZs of the C$_{60}$ monolayer and of Cu(111) are schematically drawn in Fig. 2(a). Based on the above discussion, this system is expected to be an AG with a lattice constant of 10 \AA. Figure 2(b) shows the ARPES intensity of the Fermi surface. The Shockley surface state of Cu(111) almost disappears because of the coverage of the C$_{60}$ molecules. Instead, a dot-like spectral weight can be seen at each $K$ ($K^\prime$) point. Such features do not exist on pristine Cu(111) because pristine Cu(111) only exhibits Shockley surface states at the BZ centre and bulk $sp$ bands at the BZ boundary. At deeper binding energies, the dots become closed circles and their size increases as the binding energies increase, as shown in Fig. 2(c)--2(e).

\begin{figure*}[htb]
\centering
\includegraphics[width=17 cm]{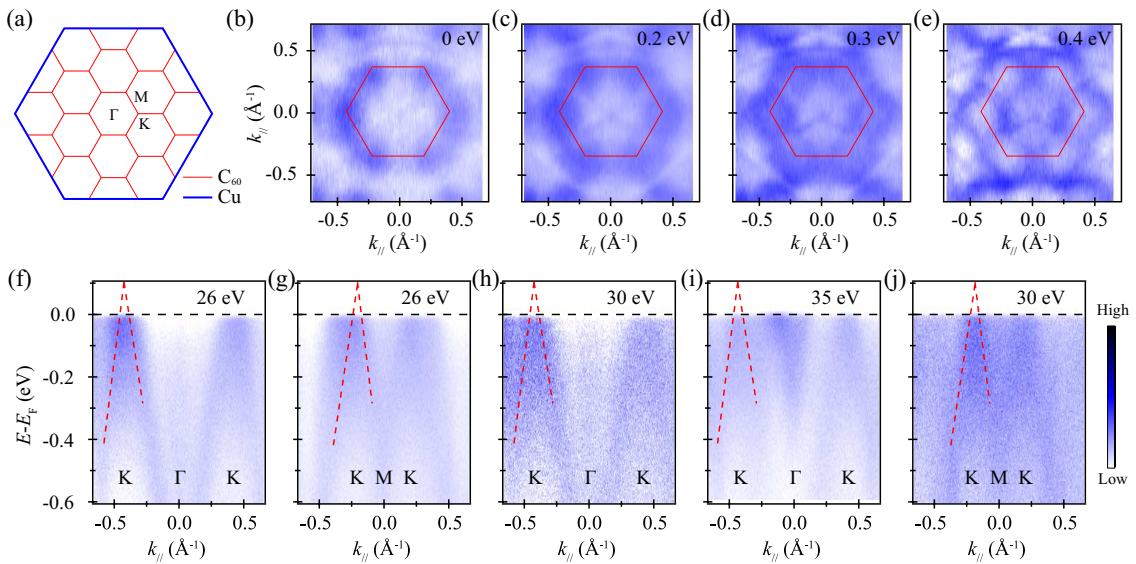}
\caption{{\bf ARPES measurements of the C$_{60}$ monolayer on Cu(111).} (a) Schematic of the BZs of the C$_{60}$ monolayer (red) and Cu(111) (blue). C$_{60}$ forms a 4$\times$4 superstructure with respect to the 1$\times$1 lattice of Cu(111). (b--e) ARPES intensity maps of the constant energy contours at different binding energies: (b) 0, (c) 0.2, (d) 0.3, and (e) 0.4 eV. Red hexagons indicate the BZs of the monolayer C$_{60}$. (f, g) ARPES intensity plots along the (f) $K$--$\Gamma$--$K$ and (g) $K$--$M$--$K$ directions. The incident photon energy is 26 eV. (h, i) ARPES intensity plots of the band structures along the $K$--$\Gamma$--$K$ direction measured with the photon energy of (h) 30 and (i) 35 eV. (j) ARPES intensity plot along the $K$--$M$--$K$ direction measured with the photon energy of 30 eV. Red dotted lines indicate the Dirac cone at the $K$ point of the BZ and are guides for the eye.}
\end{figure*}

Figure 2(f) shows the ARPES spectra along the $\Gamma$--$K$ direction. One can observe linearly dispersing bands at the $K$ point, as indicated by the red dashed lines (Fig. 1(f)). The fitted crossing point is located approximately 0.1 eV above the Fermi level. Along the $K$--$M$--$K$ direction, the dispersion of the bands is also linear, as shown in Fig. 1(g). Combined with the evolution of the constant energy contours, we can conclude that there is a Dirac cone at each $K$ point. The Dirac point is located above the Fermi level, and therefore the upper portion of the Dirac cone is inaccessible by conventional ARPES. The upward shift of the Dirac point may originate from the high electron affinity of the C$_{60}$ molecules, which leads to a significant charge transfer from the substrate to the C$_{60}$ molecules \cite{ModestiS1993}. The Fermi velocity along the $\Gamma$--$K$ direction is approximately 4$\times$10$^5$ m/s, which is slightly smaller than that of natural graphene. In addition, the Dirac bands do not disperse with different photon energies, as shown in Figs. 2(h)--2(j), which agrees with their 2D character. Therefore, the ARPES results confirm that the C$_{60}$/Cu(111) system is an AG with Dirac cones at the $K$ ($K^\prime$) points of the BZ.

We then performed model analysis to understand the physics of the AG system. We note that the Cu(111) surface hosts Shockley-type surface states inside an inverted energy gap at the centre of the BZ, and is thus a model 2DEG system. This is the reason for the success in the construction of AG by atomic/molecular manipulation \cite{GomesKK2012,WangS2014}. The Hamiltonian of a 2DEG can be described as:

\begin{figure}[htb]
\centering
\includegraphics[width=8.5 cm]{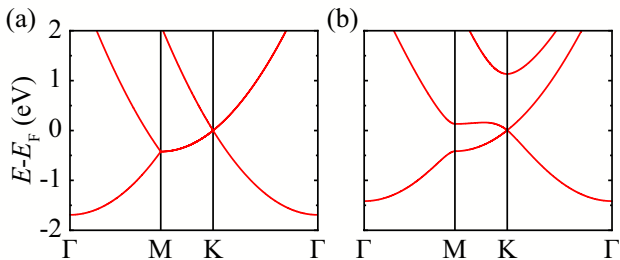}
\caption{{\bf Model analysis of artificial graphene.} (a, b) Calculated band structures for (a) U$_0$=0 eV and (b) U$_0$=0.4 eV.}
\end{figure}

\begin{equation}
H_0=-\frac{\hbar^2\nabla^2}{2m^*},
\end{equation}

\noindent where $m^*$ is the electron effective mass, $\hbar$ the reduced Planck constant, and $\nabla$ the vector differential operator. For the Cu(111) surface, $m^*$ is approximately 0.38$m_e$. The eigenvalues and eigenstates can be determined using $E(k)$=$\hbar^2k^2 $/$2m^*$ and $\psi$=e$^{ikr}$/$\sqrt{\Omega}$, where $\Omega$ is the volume of the primitive cell, $k$ the electron momentum, and $r$ the electron position. To simplify the analysis, we used a periodic potential $U(r)$ to simulate the effects of the C$_{60}$ monolayer. The Hamiltonian of the modulated 2DEG can be given as

\begin{equation}
H=H_0+U(r)=-\frac{\hbar^2\nabla^2}{2m^*}+\sum_{G_{\alpha}}U_{G_{\alpha}}e^{iG_\alpha, \cdot r}
\end{equation}

\noindent where $G_\alpha$ ($\alpha$=1,2...6) are the six nearest reciprocal lattice vectors. To simplify the analysis, the C$_{60}$ molecule can be treated as an isotropic sphere.  Therefore, the C$_{60}$/Cu(111) system possesses $C_6$ symmetry and all $U_{G_{\alpha}}$ should be equal; $i.e.$, $U_{G_1}=U_{G_2}=U_{G_3}=U_{G_4}=U_{G_5}=U_{G_6}=U_0$. As a result, the band structures of the system can be solved numerically. When the interaction between C$_{60}$ and Cu(111) is turned off, the parabolic bands of 2DEG are simply folded into the BZ of the C$_{60}$ superstructure, as shown in Fig. 3(a). With a proper $U_0$ value such as 0.4 eV, the Dirac cone at the $K$ ($K^\prime$) point emerges and resembles that of natural graphene, as shown in Fig. 3(b). The calculated Fermi velocity along the $\Gamma$--$K$ direction is approximately 6.27$\times$10$^5$ m/s, which is close to the experimental value (4$\times$10$^5$ m/s). It should be noted that the calculated Fermi velocity of the Dirac cone is insensitive to the variation of U$_0$ (See Supplemental Material, Supplemental Equations E1--E8 and Fig. S2 for details \cite{SM}).

\begin{figure*}[htb]
\centering
\includegraphics[width=14 cm]{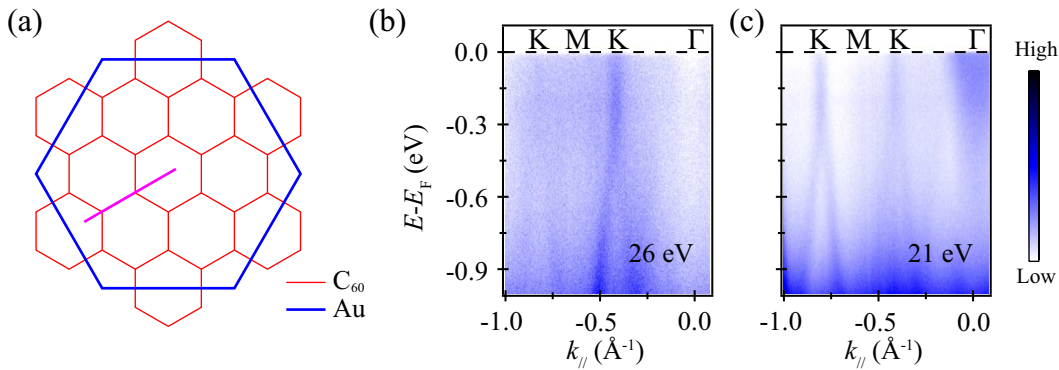}
\caption{{\bf ARPES measurements of the C$_{60}$ monolayer on Au(111).} (a) Schematic of the BZs of the C$_{60}$ monolayer (red) and Au(111) (blue). The pink line indicates the momentum direction of the cuts in (b) and (c). (b, c) ARPES intensity plots along the $\Gamma$--$K$--$M$--$K$ direction using incident photon energies of (b) 26 and (c) 21 eV. The Dirac cone at the $K$ point can be clearly observed. The Dirac point is located approximately 0.2 eV below the Fermi level.}
\end{figure*}

The success of realizing AG in the C$_{60}$/Cu(111) system provides a universal method for the realization of AG using other noble metal substrates. Another prototypical 2DEG system is Au(111), on which C$_{60}$ molecules can also form ordered monolayer structures \cite{AltmanEI1992,WangLL2004,HamadaI2011}. Because of the larger lattice constant of Au(111) ($a\rm_{Au}$=2.9 \AA) compared with that of Cu(111) ($a\rm_{Cu}$=2.5 \AA), the C$_{60}$ molecules form a 2$\sqrt3\times$2$\sqrt3$ superstructure with respect to the 1$\times$1 lattice of Au(111). This different superstructure provides further evidence for the universality of this method. As expected, our ARPES results show the existence of Dirac cones at each $K$ ($K^\prime$) point of the BZ, as shown in Fig. 4(b) and 4(c). The Dirac bands do not disperse with varying photon energy, which again agrees with their 2D character. In addition, the Dirac point of the C$_{60}$/Au(111) system is located approximately 0.2 eV below the Fermi level, and therefore the upper Dirac cone can be observed by our ARPES measurements. The higher binding energy of the Dirac point in the C$_{60}$/Au(111) system may originate from the smaller charge transfer from the substrate to C$_{60}$. This is reasonable because the electron affinity of gold is higher than that of copper. The Fermi velocity along the $\Gamma$--K direction is approximately 1.2$\times$10$^6$ m/s, which is slightly higher than that of natural graphene.

Finally, we discuss possible device applications based on these AG systems. First, Cu and Au are commonly used conducting materials, such as electrodes. After depositing a monolayer C$_{60}$, the emergence of 2D Dirac fermions enable the realization of Klein tunneling \cite{KatsnelsonMI2006,YoungAF2009,GutierrezC2016}, that is, the electrons pass unimpeded through potential barriers. This is crucial component in fabricating low-dissipation quantum devices. Second, noble metals, including Au and Cu, are well-known plasmonic materials with strong visible light response \cite{JainPK2007}. The emergence of 2D Dirac fermions on noble metal surfaces indicates the possibility of realizing Dirac plasmon \cite{LvJ2011,ChenJ2012} in the visible light region, which can be widely used in optoelectronic devices. Notably, plasmonic devices with visible light response is difficult to realize using natural graphene.

All of our results support the realization of AG in the self-assembled C$_{60}$ monolayer on noble metal surfaces. The Dirac cones at the $K$ ($K^\prime$) points of the BZ are directly observed by our ARPES measurements. The realization of AG in molecular self-assembled monolayers offers new opportunities for the fabrication of exotic molecular quantum devices. Because of the high tunability of supramolecular architectures, such molecular AG systems can also enable the investigation of Dirac fermions under various conditions, such as doping and symmetry breaking, by choosing appropriate molecules, dopants, and substrates. It should be noted that the bulk and thin film of C$_{60}$ will become a high-temperature superconductor with appropriate alkali metal doping \cite{HebardAF1991,RenMQ2020,CepekC2001}. Therefore, it is highly anticipated that the doped C$_{60}$ monolayer will host rich physics that originate from the interplay between Dirac fermions and superconductivity; these interesting properties can be probed by various experimental techniques, such as transport, scanning tunneling spectroscopy, and ARPES.

\section*{Acknowledgments}
This work was supported by the Ministry of Science and Technology of China (Grant No. 2018YFE0202700), the National Natural Science Foundation of China (Grants No. 11974391, No. 11825405, No. 1192780039, No. 11761141013), the Beijing Natural Science Foundation (Grant No. Z180007), and the Strategic Priority Research Program of the Chinese Academy of Sciences (Grants No. XDB33030100 and No. XDB30000000). The synchrotron ARPES measurements were performed with the approval of the Proposal Assessing Committee of Hiroshima Synchrotron Radiation Center (Proposal Numbers: 19AG005 and 19BG028).

S.Y. and H.Z. contributed equally to this work.

\end{document}